\documentclass[10pt,a4paper,twoside]{article}
\usepackage{epsfig}
\usepackage{baltlat6}
\usepackage{array}
\usepackage{dcolumn}
\usepackage{longtable}
\begin{document}
\ \
\vspace{0.5mm}
\setcounter{page}{125}
\vspace{8mm}

\titlehead{Baltic Astronomy, vol.\,17, 125--142, 2008}

\titleb{2MASS TWO-COLOR INTERSTELLAR REDDENING LINE IN\\ THE DIRECTION
OF THE NORTH AMERICA AND PELICAN\\ NEBULAE AND THE CYG OB2 ASSOCIATION}

\begin{authorl}
\authorb{V. Strai\v zys}{1},
\authorb{C. J. Corbally}{2} and
\authorb{V. Laugalys}{1}
\end{authorl}

\begin{addressl}
\addressb{1}{Institute of Theoretical Physics and Astronomy, Vilnius
University,\\  Go\v{s}tauto 12, Vilnius LT-01108, Lithuania;
straizys@itpa.lt; vygandas@itpa.lt}

\addressb{2}{Vatican Observatory Research Group, Steward Observatory,
Tucson, Arizona 85721, U.S.A.; corbally@as.arizona.edu}

\end{addressl}

\submitb{Received 2008 June 20; accepted 2008 June 30}

\begin{summary} The slope of the interstellar reddening line in the
$J$--$H$ vs.  $H$--$K_s$ diagram of the 2MASS survey in the direction of
the North America and Pelican nebulae, the L\,935 dust cloud and the Cyg
OB2 association is determined.  The MK types were either classified by
C. J. Corbally or collected from the literature.  The ratio
$E_{J-H}/E_{H-K_s}$ = 2.0 is obtained by taking the average for the four
groups of spectral classes:  O3--B1, B2--B6, B7--B9.5 and red clump
giants.  The obtained ratio is among the largest values of
$E_{J-H}/E_{H-K_s}$ determined till now.  \end{summary}


\resthead{2MASS two-color interstellar reddening line in Cygnus}
{V. Strai\v zys, C. J. Corbally, V. Laugalys}

\sectionb{1}{INTRODUCTION}

The 2MASS survey presents all-sky photometry in the near-infrared {\it
J, H} and $K_s$ passbands which is useful for a variety of
investigations of the Galaxy.  The system has been successfully used for
study of star-forming regions and search for young stellar objects
(YSOs), for investigation of the interstellar extinction in
dust/molecular clouds, large-scale distribution of interstellar dust,
the interstellar reddening law in the infrared, spectral energy
distributions, etc.

Recently, the interstellar reddening law in the infrared wavelengths was
studied by Fitzpatrick \& Massa (2005, 2007), Indebetouw et al.  (2005)
and Flaherty et al.  (2007) applying the 2MASS data alone or joining
them with the {\it Spitzer} results at longer wavelengths.  In some
papers regional values of the color excess ratio $E_{J-H}/E_{H-K_s}$
were
investigated (Indebetouw et al. 2005; Nishiyama et al. 2006; Lombardi et
al. 2006; Djupvik et al. 2006; Naoi et al. 2006; Rom\'an-Z\'uniga et al.
2007).  In most cases this ratio was evaluated from the statistical
distribution of red giants with various reddenings in the $J$--$H$ vs.
$H$--$K_s$ diagram.  The red clump giants of early K spectral subclasses
are the most abundant population in 2MASS near the Galactic plane, since
they are very bright in the near infrared and, despite the interstellar
extinction, have been observed at large distances -- up to the Galactic
central bulge and the disk edges in the direction of the 2nd and 3rd
quadrants.

In the listed papers some regional variations of the $E_{J-H}/E_{H-K_s}$
ratio or the interstellar reddening law in the infrared wavelengths have
been noted.  Consequently, before applying the $J$--$H$ vs.  $H$--$K_s$
diagram for star classification or other tasks, it is important to
investigate the color-excess ratio in each specific area.

To our knowledge, the ratio $E_{J-H}/E_{H-K_s}$ in the area of the North
America and Pelican nebulae (hereafter NAP), including the L\,935 dust
cloud, separating the nebulae, has not been determined till now.
Cambr\'esy et al.  (2002), Comer\'on et al.  (2002) and Comer\'on \&
Pasquali
(2005) in their studies, based on the 2MASS data in Cygnus, have applied
a value of 1.70 which follows from the Rieke \& Lebofsky (1985)
interstellar extinction law.  However, this ratio corresponds to the
Arizona {\it J,H,K} system, which is slightly different from the 2MASS
system.  On the other hand, this extinction law was obtained by using
heavily reddened stars located in various directions (five stars in the
Galactic center direction, $o$ Sco and Cyg OB2 No.\,12).  Among them,
Cyg OB2 No.\,12 is an emission-line B supergiant with a number of
peculiarities, see discussion in Section 3.

We decided to investigate the ratio $E_{J-H}/E_{H-K_s}$ in the NAP
direction applying the classical method which determines slopes of
reddening lines plotted for stars in narrow intervals of spectral
classes.  For this aim the most suitable are early-type stars of
spectral classes O and B since these types of stars are sufficiently
luminous and apparently bright to be accessible for spectral
classification and sufficiently distant to be considerably reddened.
A similar method has been successfully used by He et al.  (1995) for
reddened O--B stars in the southern Milky Way.

\sectionb{2}{THE LIST OF O--B STARS}

The starting step in composing the list of early-type stars in NAP,
classified in the MK system, was the search of the Simbad database for
the ALS stars (Reed 1998, 2005) in the area 3\degr\,$\times$\,3\degr\
with the center at J2000: 20$^{\rm h}$\,56$^{\rm m}$, +44\degr.  The
next step was the check of the
catalogs of stars measured and classified in two dimensions using the
{\it Vilnius} seven-color photometric system (Strai\v{z}ys et al. 1989,
1993; Laugalys \& Strai\v{z}ys 2002; Laugalys et al. 2006a,b, 2007).  To
verify the quality of photometric classifications, 37 B-stars from the
Vilnius lists were classified by one of us (C.J.C.) using  blue
grating spectra of 2.8 \AA\ resolution obtained with the Boller and
Chivens spectrograph on the 2.3 m telescope of Steward Observatory at
Kitt Peak.  Part of these classifications were published in Strai\v{z}ys
et al.  (1999), the spectral types of the remaining 14 are given in
Table 1 of the present paper.

Since the coincidence between spectroscopic and photometric spectral
types was quite good, we have added to the list 29 B-type stars in the
cluster NGC 6997 and Collinder 428 areas with reliable two-dimensional
classifications obtained using the {\it Vilnius} seven-color photometric
system (Laugalys et al. 2006a, 2007); in Table 1 their spectral classes
are marked by lower-case letters.

However, for the selected B-stars in the region of the North America and
Pelican nebulae the largest values of $E_{H-K_s}$ are about 0.25 which
correspond to $A_V$\,$\approx$\,4 mag.  To have a longer reddening line,
we added O--B stars from the\break

%%%%%%%%%%%%%%%%%%%%%%%%%%%%%%  TABLE 1
\vspace{-8mm}
\begin{center}
\small
\noindent
\tabcolsep=2pt
\begin{longtable}{llccD..{-1}lD..{1.3}D..{2.3}}
\multicolumn{8}{l}{\parbox{125mm}{\baselineskip=9pt
{\normbf\ \ Table 1.}{\norm\ O--B stars in the vicinity of the North
America and Pelican nebulae and in the Cyg OB2 association.\lstrut}}}\\
\hline
\noalign{\vskip1.5mm}
\multicolumn{1}{c}{Name*}  &
\multicolumn{1}{c}{BD, HD} &
\multicolumn{1}{c}{RA\,(2000)}   &
\multicolumn{1}{c}{DEC\,(2000)}  &
\multicolumn{1}{c}{\huad $V$}  &
\multicolumn{1}{c}{Sp\hhuad\hhuad} &
\multicolumn{1}{c}{$J$--$H$} &
\multicolumn{1}{c}{$H$--$K_s$}  \\
\noalign{\vskip1.5mm}
\noalign{\vskip1.5mm}
\hline
\noalign{\vskip1mm}
\endfirsthead
\multicolumn{8}{l}{{\normbf\ \ Table 1.}{\norm\ Continued \lstrut}}\\
\hline
\noalign{\vskip1.5mm}
\multicolumn{1}{c}{Name*}  &
\multicolumn{1}{c}{BD, HD} &
\multicolumn{1}{c}{RA\,(2000)}   &
\multicolumn{1}{c}{DEC\,(2000)}  &
\multicolumn{1}{c}{\huad $V$}  &
\multicolumn{1}{c}{Sp\hhuad\hhuad} &
\multicolumn{1}{c}{$J$--$H$} &
\multicolumn{1}{c}{$H$--$K_s$}  \\
\noalign{\vskip1.5mm}
\hline
\noalign{\vskip1mm}
\endhead
%\hline
\endfoot
ALS\,17661  & BD+45 3239   &  20 40 58.94  &  45 46 59.2 &   8.71  &  B6\,III       &   0.078  &   0.050  \\
ALS\,11494  & BD+42 3835   &  20 42 06.86  &  43 11 03.7 &   9.20  &  O9           &   0.245  &   0.121  \\
ALS\,11500  & BD+45 3246   &  20 42 31.01  &  45 54 04.6 &   9.67  &  B1\,Vn        &  -0.042  &   0.054  \\
ALS\,11525  & BD+45 3260   &  20 45 35.28  &  46 21 02.1 &   9.06  &  O9\,V         &   0.098  &   0.084  \\
ALS\,17656  & BD+45 3264   &  20 45 52.34  &  45 46 47.5 &   9.35  &  B8\,III       &  -0.053  &  -0.037  \\
ALS\,17618  & BD+41 3884   &  20 47 09.92  &  42 24 35.4 &   7.42  &  B9\,III       &  -0.095  &  -0.005  \\
ALS\,17619  & HD 198414    &  20 48 26.34  &  45 27 07.6 &   7.68  &  B7\,III       &  -0.029  &  -0.002  \\
ALS\,11565  & BD+45 3290   &  20 48 54.99  &  45 37 27.2 &   8.57  &  B1\,III       &   0.032  &   0.042  \\
ALS\,11566  & BD+45 3291   &  20 48 56.29  &  46 06 50.9 &   4.83  &  B3\,Ia        &   0.133  &   0.103  \\
ALS\,11568  & BD+44 3594   &  20 49 11.59  &  45 24 39.8 &   9.78  &  B1\,V:npe     &   0.217  &   0.330  \\
ALS\,11576  & BD+45 3295   &  20 49 44.20  &  46 00 33.8 &  10.31  &  B3\,II        &   0.087  &  -0.029  \\
SMV89-131  &              &  20 51 40.00  &  44 00 02.7 &  10.78  &  B2\,V*         &   0.250  &   0.084  \\
ALS\,11593  &              &  20 51 44.07  &  46 01 48.3 &  10.54  &  B0.5\,Vn       &   0.091  &   0.093  \\
ALS\,17657  & HD 198915    &  20 51 57.35  &  46 44 05.2 &   7.50  &  B6\,IV        &  -0.111  &  -0.031  \\
SMV89-140  &              &  20 52 07.65  &  44 03 44.4 &  10.52  &  B3\,V*         &   0.203  &   0.075  \\
ALS\,11597  & HD 198931    &  20 52 09.72  &  44 26 04.6 &   8.72  &  B1\,Ve*        &   0.266  &   0.328  \\
LS02-40    &              &  20 52 23.75  &  44 36 15.1 &  12.09  &  B2.5\,Vnn**    &   0.337  &   0.146  \\
SMV89-148  & BD+43 3751   &  20 52 33.36  &  44 11 42.1 &   9.66  &  B2\,Vp*        &   0.053  &   0.123  \\
SKV2-44    &              &  20 52 46.71  &  43 55 43.5 &  11.02  &  B7\,V*         &   0.178  &   0.058  \\
ALS\,11599  & HD 199021    &  20 52 53.21  &  42 36 27.9 &   8.43  &  B0\,V         &   0.113  &   0.068  \\
SKV2-52    &              &  20 53 04.97  &  43 37 13.2 &  11.18  &  B2\,Vn*        &   0.301  &   0.108  \\
SMV89-159*  & HD 199081    &  20 53 14.75  &  44 23 14.1 &   4.74  &  B5\,V     &   0.148  &   0.002  \\
SKV2-57    &              &  20 53 15.91  &  43 47 57.7 &  12.49  &  B1\,V*         &   0.360  &   0.223  \\
SMV89-163  &              &  20 53 34.19  &  44 09 07.0 &  11.45  &  B3\,V*         &   0.270  &   0.087  \\
SMV89-166  & BD+42 3897   &  20 53 39.23  &  42 42 05.4 &   8.68  &  B8.5\,V       &  -0.045  &   0.015  \\
ALS\,17620*  & HD 199206    &  20 54 05.94  &  45 06 36.5 &   7.67  &  B8\,II    &  -0.031  &   0.004  \\
ALS\,17658  & BD+46 3097   &  20 54 20.74  &  46 42 40.8 &   9.12  &  B9.5\,III     &   0.023  &   0.055  \\
ALS\,17622*  & HD 199312    &  20 54 45.30  &  45 08 10.6 &   7.59  &  B8\,IV   &  -0.045  &  -0.028  \\
ALS\,11618  & BD+44 3627   &  20 54 47.47  &  44 50 46.5 &   9.85  &  B2\,III*       &   0.200  &   0.131  \\
LS02-209   &              &  20 54 58.87  &  44 54 55.6 &  11.58  &  B1\,V**         &   0.183  &   0.146  \\
SMV89-177  & BD+44 3629   &  20 55 04.08  &  44 45 34.1 &  10.06  &  B8\,V*         &   0.136  &   0.078  \\
LS02-217   &              &  20 55 04.40  &  45 20 48.8 &  12.36  &  B4\,V**         &   0.079  &   0.030  \\
LS02-222   &              &  20 55 07.29  &  44 35 49.8 &  12.24  &  B0.5\,III**     &   0.385  &   0.154  \\
LS02-229   &              &  20 55 10.27  &  44 42 46.4 &  12.36  &  B8\,III-IVp**   &   0.393  &   0.229  \\
LS02-230   &              &  20 55 10.31  &  45 03 03.1 &  11.84  &  B0.5:\,Ve**     &   0.314  &   0.278  \\
SMV89-184  & BD+42 3909   &  20 55 33.07  &  43 32 55.6 &   8.65  &  B9.5\,V       &  -0.048  &   0.019  \\
LS02-267   &              &  20 55 35.60  &  45 09 00.9 &  11.99  &  B1\,V**         &   0.078  &   0.043  \\
SMV89-185*  & BD+44 3636   &  20 55 48.10  &  44 47 44.7 &  10.39  &  B9\,V     &   0.099  &   0.086  \\
SMV89-188*  & HD 199479    &  20 55 59.01  &  44 22 26.2 &   6.81  &  B9\,V     &  -0.026  &   0.004  \\
CP05\,4    &              &  20 55 51.25  &  43 52 24.5 &  13.24  &  O5\,V         &   0.849  &   0.466  \\
SMV89-190  &              &  20 56 02.96  &  45 21 24.0 &  11.27  &  B5\,IV*        &   0.009  &   0.082  \\
LSV06-1-133 &             &  20 56 07.11  &  44 42 23.1 &  12.81  &  b4\,V         &   0.236  &   0.091  \\
LSV06-1-158 &             &  20 56 11.42  &  44 27 56.0 &  15.32  &  b7\,III       &   0.310  &   0.186  \\
LSV06-1-175 &             &  20 56 13.76  &  44 38 39.9 &  16.24  &  b9\,V         &   0.445  &   0.144  \\
SMV89-194   & BD+44 3637  &  20 56 18.02  &  44 46 47.8 &   9.68  &  B8\,V         &  -0.096  &   0.009  \\
LSV06-1-203 &             &  20 56 19.27  &  44 40 34.5 &  14.92  &  b9\,V         &   0.329  &   0.184  \\
ALS\,11628  & BD+42 3914  &  20 56 24.09  &  43 07 46.5 &   8.43  &  B0\,III:      &   0.287  &   0.162  \\
SMV89-197  &              &  20 56 24.61  &  44 39 21.5 &  11.13  &  B5\,V*         &   0.136  &   0.034  \\
LSV06-1-242 &             &  20 56 25.66  &  44 38 55.8 &  15.72  &  b7\,III       &   0.308  &   0.191  \\
LSV06-1-295 &             &  20 56 32.52  &  44 45 27.4 &  16.21  &  b9\,V         &   0.350  &   0.208  \\
ALS\,11633*  & HD 199579  &  20 56 34.78  &  44 55 29.0 &   5.96  &  O6.5\,III    &  -0.035  &  -0.026  \\
LS02-691   &              &  20 56 37.18  &  43 55 05.5 &  13.88   &  B2\,Ib**        &   0.483  &   0.249  \\
ALS\,11636  & BD+45 3339  &  20 56 39.23  &  46 21 20.7 &   9.93  &  B1\,IV        &   0.118  &   0.007  \\
LSV06-1-356 &             &  20 56 42.28  &  44 39 16.8 &  14.00  &  b9\,III       &   0.321  &   0.138  \\
LSV06-22-35 &             &  20 56 43.74  &  43 53 25.3 &  17.62  &  b8\,V:        &   0.520  &   0.250  \\
LSV06-1-366 &             &  20 56 43.82  &  44 25 56.5 &  14.98  &  b9\,IV-V      &   0.375  &   0.176  \\
LSV06-22-47 &             &  20 56 50.09  &  43 56 23.8 &  15.46  &  b9.5\,IV      &   0.437  &   0.212  \\
LSV06-1-427 &             &  20 56 52.50  &  44 35 53.0 &  13.75  &  b5\,V         &   0.264  &   0.111  \\
ALS\,11643   & BD+45 3341  &  20 57 02.68  &  46 32 44.7 &   8.73  &  B1\,II        &   0.105  &   0.019  \\
LSV06-1-497 &             &  20 57 02.79  &  44 36 45.1 &  14.91  &  b9.5\,III     &   0.340  &   0.122  \\
LS02-401   &              &  20 57 04.12  &  45 12 53.3 &  12.16  &  B9\,III**       &   0.145  &   0.088  \\
LSV06-1-525 &             &  20 57 06.35  &  44 31 29.8 &  15.86  &  b9\,V         &   0.318  &   0.162  \\
LS02-463   &              &  20 57 49.37  &  44 51 27.2 &  12.32  &  B5\,V He(e)**   &   0.163  &   0.064  \\
LS02-476   &              &  20 57 54.38  &  44 31 38.3 &  12.49  &  B4\,Vp(Si)**    &   0.218  &   0.176  \\
ALS\,11651  & BD+44 3655   &  20 58 25.52  &  45 08 59.1 &   9.24  &  B1\,IV*        &   0.074  &   0.060  \\
ALS\,16465  & BD+41 3949   &  20 58 30.95  &  41 56 23.7 &   6.16  &  B7\,III       &  -0.052  &  -0.030  \\
LS02-537   &              &  20 58 36.57  &  45 05 02.6 &  10.92  &  B1\,Ve**        &   0.282  &   0.311  \\
SKV2-198   &              &  20 59 01.57  &  42 55 42.5 &  11.34  &  B9.5\,IV-V*    &   0.238  &   0.134  \\
LS02-584   &              &  20 59 04.72  &  44 06 41.2 &  12.90  &  B4\,IV**        &   0.419  &   0.261  \\
SMV89-226  &              &  20 59 14.49  &  44 46 57.3 &  10.33  &  B4\,IIIn*      &   0.172  &   0.099  \\
LS02-608   &              &  20 59 18.52  &  45 29 50.0 &  11.18  &  B0.5\,Vne**     &   0.176  &   0.214  \\
ALS\,19944  & HD 200030    &  20 59 24.62  &  42 19 28.1 &   6.48  &  B8\,III       &  -0.084  &  -0.018  \\
SMV89-229  & HD 200042    &  20 59 33.11  &  43 03 51.5 &   8.01  &  B7\,III       &  -0.025  &   0.028  \\
SMV89-230  &              &  20 59 30.70  &  45 17 19.1 &  11.08  &  B4\,III*       &   0.116  &   0.061  \\
SMV89-236  & BD+44 3664   &  20 59 55.99  &  45 20 13.0 &  10.19  &  B1\,Vn*        &   0.151  &   0.088  \\
SMV89-239  & BD+44 3666   &  21 00 05.18  &  45 02 49.3 &  10.18  &  B1\,Ve*        &   0.207  &   0.264  \\
SMV89-242  & HD 200178    &  21 00 28.73  &  43 33 40.4 &   8.36  &  B9\,V*         &  -0.015  &  -0.017  \\
ALS\,11675  & BD+45 3360   &  21 00 34.21  &  46 14 49.9 &  10.00  &  B3\,V         &   0.094  &   0.028  \\
ALS\,17624  & BD+46 3141   &  21 00 49.86  &  46 34 42.9 &   7.26  &  B5\,V         &  -0.055  &  -0.044  \\
SMV89-245  & BD+42 3937   &  21 01 00.92  &  42 46 31.4 &   9.34  &  B8\,III-IV*    &   0.183  &   0.010  \\
ALS\,11678  & BD+45 3364   &  21 01 10.93  &  46 09 20.8 &   5.40  &  B1\,V         &  -0.092  &  -0.041  \\
SMV89-248* & HD 200311    &  21 01 14.32  &  43 43 18.4 &   7.66  &  B8p         &  -0.080  &  -0.010  \\
SMV89-249  &              &  21 01 24.34  &  44 10 01.1 &  11.18  &  B5\,III*       &   0.203  &   0.087  \\
ALS\,11682  &              &  21 02 12.58  &  46 12 37.8 &  10.19  &  B8\,II        &   0.028  &   0.044  \\
LSV07-114  &              &  21 02 41.03  &  44 31 54.0 &  13.12  &  b8\,IV        &   0.194  &   0.063  \\
LSV07-217  &              &  21 02 52.06  &  44 35 33.2 &  11.39  &  b5.5\,III     &   0.123  &   0.056  \\
LSV07-198  &              &  21 02 50.05  &  44 29 48.5 &  12.96  &  b9\,IV-V      &   0.187  &   0.070  \\
LSV07-201  &              &  21 02 50.42  &  44 37 14.7 &  12.36  &  b8\,IV-V      &   0.182  &   0.019  \\
LSV07-355  &              &  21 03 03.81  &  44 34 37.2 &  13.31  &  b7\,IV        &   0.126  &   0.046  \\
LSV07-464  &              &  21 03 13.95  &  44 45 03.7 &  13.54  &  b6\,III       &   0.163  &   0.110  \\
LSV07-357  &              &  21 03 04.38  &  44 34 37.6 &  13.97  &  b7\,V         &   0.096  &   0.110  \\
LSV07-498  &              &  21 03 16.88  &  44 32 59.4 &  14.23  &  b6\,V         &   0.116  &   0.024  \\
LSV07-363  &              &  21 03 05.26  &  44 37 14.2 &  14.30  &  b7\,IV        &   0.151  &   0.108  \\
ALS\,17978  & BD+44 3685   &  21 03 38.45  &  45 22 04.6 &   7.86  &  B8\,II        &  -0.076  &   0.009  \\
ALS\,11699  & BD+45 3384   &  21 03 53.80  &  46 19 49.9 &   7.81  &  B1\,IV:p      &  -0.021  &  -0.039  \\
LSV07-827   &              &  21 03 58.01  &  44 35 32.8 &  14.40  &  b9\,III-IV    &   0.179  &   0.070  \\
LSV07-479   &              &  21 03 15.67  &  44 41 21.2 &  15.79  &  b6\,IV        &   0.161  &   0.154  \\
LSV07-665   &              &  21 03 34.33  &  44 45 24.0 &  15.83  &  b9\,Vp?       &   0.171  &   0.157  \\
LSV07-447   &              &  21 03 12.27  &  44 37 37.7 &  15.98  &  b8\,IV        &   0.157  &   0.176  \\
LSV07-438   &              &  21 03 11.06  &  44 46 43.0 &  16.17  &  b3\,V         &   0.200  &   0.137  \\
LSV07-690   &              &  21 03 36.08  &  44 26 40.4 &  16.30  &  b7\,V         &   0.408  &   0.173  \\
ALS\,17987  & BD+45 3387   &  21 04 18.21  &  46 31 52.8 &   8.61  &  B8\,III       &  -0.008  &   0.003  \\
ALS\,11718  & BD+45 3406   &  21 06 32.43  &  45 51 31.0 &   9.52  &  B1\,Iab       &   0.134  &   0.112  \\
ALS\,17980  & BD+44 3710   &  21 07 14.94  &  44 40 26.7 &   7.38  &  B8\,III       &
\multicolumn{1}{c}{--} &
\multicolumn{1}{c}{--}  \\
\noalign{\vskip2mm}
\noalign{Cyg OB2 association: brightest stars}
\noalign{\vskip2mm}
VI Cyg 1        &              &  20 31 10.55  &  41 31 53.5 &  11.06  &  O9\,V         &   0.412  &   0.191  \\
VI Cyg 2        &              &  20 31 22.04  &  41 31 28.4 &  10.61  &  B1\,Ib:       &   0.325  &   0.122  \\
VI Cyg 3        & BD+40 4212   &  20 31 37.50  &  41 13 21.0 &  10.35  &  O9:           &   0.497  &   0.253  \\
VI Cyg 4        & BD+40 4219   &  20 32 13.83  &  41 27 12.0 &  10.07  &  O7\,III       &   0.334  &   0.143  \\
VI Cyg 5*       & BD+40 4220   &  20 32 22.43  &  41 18 19.1 &   9.21  &  O7e           &   0.442  &   0.406  \\
VI Cyg 12       &              &  20 32 40.96  &  41 14 29.2 &  10.40  &  B5\,Iab:      &   1.155  &   0.808  \\
VI Cyg 6        &              &  20 32 45.46  &  41 25 37.4 &  10.65  &  O8\,V:        &   0.336  &   0.196  \\
VI Cyg 9        &              &  20 33 10.75  &  41 15 08.2 &  10.78  &  O5\,Ie        &   0.571  &   0.327  \\
VI Cyg 7        &              &  20 33 14.11  &  41 20 21.8 &  10.50  &  O3\,I         &   0.430  &   0.207  \\
VI Cyg 8B       &              &  20 33 14.76  &  41 18 41.6 &  10.31  &  O8            &   0.447  &   0.192  \\
VI Cyg 8A       & BD+40 4227   &  20 33 15.08  &  41 18 50.5 &   8.99  &  O5.5\,I       &   0.402  &   0.218  \\
VI Cyg 8D       &              &  20 33 16.34  &  41 19 01.8 &  12.02  &  O8\,V         &   0.418  &   0.185  \\
VI Cyg 8C       &              &  20 33 17.99  &  41 18 31.1 &  10.08  &  O5\,III       &   0.373  &   0.213  \\
VI Cyg 10       & BD+41 3804   &  20 33 46.10  &  41 33 01.1 &   9.89  &  O9.5\,Ia      &   0.455  &   0.257  \\
VI Cyg 11       & BD+41 3807   &  20 34 08.50  &  41 36 59.2 &  10.08  &  O5\,If        &   0.424  &   0.236  \\
\noalign{\vskip2mm}
\noalign{Cyg OB2 association: Massey \& Thompson (1991) stars}
\noalign{\vskip2mm}
MT91-299   &              &  20 32 38.57  &  41 25 13.7 &  10.84  &  O7.5\,V       &   0.276  &   0.202  \\
MT91-556   &              &  20 33 30.78  &  41 15 22.6 &  11.01  &  B1\,Ib        &   0.602  &   0.349  \\
MT91-601   &              &  20 33 39.10  &  41 19 25.8 &  11.07  &  O9.5\,III     &   0.485  &   0.263  \\
MT91-258   &              &  20 32 27.66  &  41 26 22.0 &  11.10  &  O8\,V         &   0.342  &   0.172  \\
MT91-259   &              &  20 32 27.74  &  41 28 52.2 &  11.42  &  B0.5\,V       &   0.296  &   0.129  \\
MT91-227   &              &  20 32 16.56  &  41 25 35.7 &  11.47  &  O9\,V         &   0.325  &   0.204  \\
MT91-145   &              &  20 31 49.65  &  41 28 26.5 &  11.52  &  O9.5\,V       &   0.306  &   0.134  \\
MT91-417   &              &  20 33 08.79  &  41 13 18.2 &  11.55  &  O4\,III(f)    &   0.570  &   0.314  \\
MT91-531   &              &  20 33 25.56  &  41 33 26.9 &  11.58  &  O8.5\,V       &   0.420  &   0.225  \\
MT91-339   &              &  20 32 50.02  &  41 23 44.6 &  11.60  &  O8.5\,V       &   0.391  &   0.206  \\
MT91-605   &              &  20 33 39.79  &  41 22 52.3 &  11.78  &  B0.5\,V       &   0.333  &   0.264  \\
MT91-642   &              &  20 33 47.83  &  41 20 41.5 &  11.78  &  B1\,III-Ib    &   0.499  &   0.278  \\
MT91-516   &              &  20 33 23.46  &  41 09 13.0 &  11.84  &  O5.5\,V(f)    &   0.645  &   0.330  \\
MT91-480   &              &  20 33 17.48  &  41 17 09.3 &  11.88  &  O7.5\,V       &   0.465  &   0.240  \\
MT91-745   &              &  20 34 13.50  &  41 35 02.7 &  11.91  &  O7\,V         &   0.402  &   0.227  \\
MT91-376   &              &  20 32 59.19  &  41 24 25.4 &  11.91  &  O8\,V         &   0.362  &   0.210  \\
MT91-473   &              &  20 33 16.34  &  41 19 01.7 &  12.02  &  O8.5\,V       &   0.418  &   0.185  \\
MT91-771   &              &  20 34 29.59  &  41 31 45.5 &  12.06  &  O7\,V         &   0.530  &   0.321  \\
MT91-485   &              &  20 33 18.03  &  41 21 36.6 &  12.06  &  O8\,V         &   0.429  &   0.202  \\
MT91-138   &              &  20 31 45.40  &  41 18 26.7 &  12.26  &  O8.5\,I       &   0.513  &   0.293  \\
MT91-793   &              &  20 34 43.58  &  41 29 04.6 &  12.29  &  B1.5\,III:    &   0.498  &   0.415  \\
MT91-696   &              &  20 33 59.52  &  41 17 35.4 &  12.32  &  O9.5\,V       &   0.394  &   0.251  \\
MT91-588   &              &  20 33 37.00  &  41 16 11.3 &  12.40  &  B0\,V         &   0.515  &   0.239  \\
MT91-470   &              &  20 33 15.71  &  41 20 17.2 &  12.50  &  O9.5\,V       &   0.398  &   0.210  \\
MT91-555   &              &  20 33 30.30  &  41 35 57.8 &  12.51  &  O8\,V         &   0.546  &   0.271  \\
MT91-174   &              &  20 31 56.94  &  41 31 47.8 &  12.55  &  B1.5\,V       &   0.388  &   0.182  \\
MT91-507   &              &  20 33 21.01  &  41 17 40.1 &  12.70  &  O8.5\,V       &   0.402  &   0.227  \\
MT91-611   &              &  20 33 40.86  &  41 30 18.9 &  12.77  &  O7\,Vp        &   0.397  &   0.252  \\
MT91-736   &              &  20 34 09.51  &  41 34 13.6 &  12.79  &  O9\,V         &   0.412  &   0.246  \\
MT91-455   &              &  20 33 13.69  &  41 13 05.7 &  12.92  &  O8\,V         &   0.475  &   0.279  \\
MT91-5     &              &  20 30 39.81  &  41 36 50.7 &  12.93  &  O6\,V(f)      &   0.524  &   0.261  \\
MT91-403   &              &  20 33 05.26  &  41 43 36.7 &  12.94  &  B1.5\,V       &   0.432  &   0.230  \\
MT91-390   &              &  20 33 02.92  &  41 17 43.1 &  12.95  &  O8\,V         &   0.553  &   0.292  \\
MT91-429   &              &  20 33 10.50  &  41 22 22.4 &  12.98  &  B0\,V         &   0.424  &   0.216  \\
MT91-70    &              &  20 31 18.33  &  41 21 21.6 &  12.99  &  O9\,V         &   0.561  &   0.300  \\
MT91-534   &              &  20 33 26.74  &  41 10 59.5 &  13.00  &  O7.5\,V       &   0.537  &   0.269  \\
MT91-292   &              &  20 32 37.02  &  41 23 05.2 &  13.08  &  B1\,V         &   0.465  &   0.221  \\
MT91-187   &              &  20 32 03.77  &  41 25 10.4 &  13.24  &  B0.5:\,V      &   0.405  &   0.235  \\
MT91-248   &              &  20 32 25.49  &  41 24 52.0 &  13.36  &  O5.5\,V       &   0.344  &   0.203  \\
MT91-575   &              &  20 33 34.33  &  41 18 11.3 &  13.41  &  B1.5\,V       &   0.679  &   0.572  \\
MT91-467   &              &  20 33 15.31  &  41 29 56.7 &  13.43  &  B1\,V         &   0.456  &   0.230  \\
MT91-378   &              &  20 32 59.64  &  41 15 14.6 &  13.49  &  B0\,V         &   0.596  &   0.307  \\
MT91-716   &              &  20 34 04.86  &  41 05 12.9 &  13.50  &  O9\,V         &   0.466  &   0.259  \\
MT91-692   &              &  20 33 59.25  &  41 05 38.0 &  13.61  &  B0:\,V        &   0.421  &   0.266  \\
MT91-448   &              &  20 33 13.26  &  41 13 28.7 &  13.61  &  O6\,V(f)      &   0.636  &   0.337  \\
\noalign{\vskip2mm}
\noalign{Cyg OB2 association: Comer\'on et al. (2002) stars}
\noalign{\vskip2mm}
CPR02-A4  &  &   20 31 36.25  &  41 22 03.2 & 15.0   &    ~~~--       &  1.149  &   0.655     \\
CPR02-A5  &  &   20 35 09.75  &  41 35 29.7 & 16.2   &    ~~~--       &  1.040  &   0.611     \\
CPR02-A6  &  &   20 32 08.33  &  40 25 06.9 & 17.5   &    ~~~--       &  1.055  &   0.548     \\
CPR02-A7  &  &   20 34 42.96  &  40 29 30.2 & 17.7   &    ~~~--       &  1.072  &   0.541     \\
CPR02-A8  &  &   20 33 41.61  &  41 47 57.0 & 15.5   &    ~~~--       &  1.165  &   0.564     \\
CPR02-A9  &  &   20 35 32.71  &  41 20 55.0 & 17.2   &    ~~~--       &  1.019  &   0.498     \\
CPR02-A10 &  &   20 34 55.11  &  40 34 44.3 & 15.6   &    ~~~--       &  0.840  &   0.451     \\
CPR02-A11 &  &   20 32 31.54  &  41 14 08.2 & 12.5   &  O7.5\,Ib-II    &  0.723  &   0.430     \\
CPR02-A12 &  &   20 33 38.21  &  40 41 06.4 & 12.1   &  B0\,Ia       &  0.734  &   0.425     \\
CPR02-A13 &  &   20 33 01.24  &  40 32 33.7 & 15.1   &    ~~~--       &  0.745  &   0.398     \\
CPR02-A14 &  &   20 31 18.99  &  42 02 55.9 & 14.3   &    ~~~--       &  0.768  &   0.402     \\
CPR02-A15 &  &   20 31 36.90  &  40 59 09.2 & 13.0   &  O7\,Ib(f)     &  0.705  &   0.397     \\
CPR02-A16 &  &   20 34 36.94  &  40 41 01.9 & 15.0   &    ~~~--       &  0.737  &   0.386     \\
CPR02-A17 &  &   20 32 35.34  &  41 14 45.4 & 14.5   &    ~~~--       &  0.644  &   0.388     \\
CPR02-A18 &  &   20 30 07.88  &  41 23 50.4 & 14.0   &  $\sim$O8\,V        &  0.658  &   0.374     \\
CPR02-A19 &  &   20 31 25.91  &  41 16 02.7 & 14.1   &    ~~~--       &  0.659  &   0.341     \\
CPR02-A20 &  &   20 33 02.92  &  40 47 25.4 & 12.0   &  O8\,IIf       &  0.619  &   0.358     \\
CPR02-A21 &  &   20 29 34.80  &  41 20 08.9 & 13.7   &    ~~~--       &  0.635  &   0.331     \\
CPR02-A22 &  &   20 33 11.29  &  40 42 33.8 & 13.4   &    ~~~--       &  0.649  &   0.330     \\
CPR02-A23 &  &   20 30 39.70  &  41 08 48.9 & 11.3   &  B0.7\,Ib      &  0.600  &   0.348     \\
CPR02-A24 &  &   20 34 44.10  &  40 51 58.4 & 12.7   &  O6.5\,III(f)  &  0.609  &   0.348     \\
CPR02-A25 &  &   20 32 38.43  &  40 40 44.5 & 13.0   &  $\sim$O8\,III       &  0.642  &   0.322     \\
CPR02-A26 &  &   20 30 57.72  &  41 09 57.5 & 13.1   &  O9.5\,V       &  0.579  &   0.316     \\
CPR02-A27 &  &   20 34 44.71  &  40 51 46.5 & 11.4   &  B0\,Ia        &  0.621  &   0.331     \\
CPR02-A28 &  &   20 34 16.04  &  41 02 19.6 & 13.4   &    ~~~--       &  0.579  &   0.313     \\
CPR02-A29 &  &   20 34 56.05  &  40 38 18.0 & 11.7   &  O9.7\,Iab     &  0.581  &   0.314     \\
CPR02-A30 &  &   20 31 22.10  &  41 12 02.9 & 13.1   &  $\sim$B2\,V         &  0.469  &   0.299     \\
CPR02-A31 &  &   20 32 39.49  &  40 52 47.5 & 13.1   &  $\sim$B0.5\,V       &  0.595  &   0.285     \\
CPR02-A32 &  &   20 32 30.33  &  40 34 33.2 & 12.1   &  O9.5\,IV      &  0.527  &   0.295     \\
CPR02-A33 &  &   20 32 34.98  &  40 52 39.0 & 13.0   &  B2.5\,V       &  0.541  &   0.289     \\
CPR02-A34 &  &   20 31 36.93  &  42 01 21.8 & 11.3   &  B0.7\,Ib      &  0.399  &   0.299     \\
CPR02-A35 &  &   20 30 55.52  &  40 54 54.1 & 12.8   &  $\sim$B0\,V   &  0.488  &   0.285     \\
CPR02-A36 &  &   20 34 58.78  &  41 36 17.4 & 11.5   &  B0\,Ibn       &  0.538  &   0.299     \\
CPR02-A37 &  &   20 36 04.51  &  40 56 12.9 & 12.3   &  O5\,Vf        &  0.600  &   0.283     \\
CPR02-A38 &  &   20 32 34.86  &  40 56 17.4 & 13.1   &  O8\,V         &  0.524  &   0.294     \\
CPR02-A39 &  &   20 32 27.34  &  40 55 18.4 & 11.9   &  B2\,V         &  0.466  &   0.277     \\
CPR02-A40 &  &   20 35 13.66  &  40 55 25.0 & 12.5   &    ~~~--       &  0.608  &   0.257     \\
CPR02-A41 &  &   20 31 08.38  &  42 02 42.2 & 12.4   &  O9.7\,II      &  0.536  &   0.269     \\
CPR02-A42 &  &   20 29 57.01  &  41 09 53.8 & 12.3   &  B0\,V         &  0.417  &   0.250     \\
CPR02-A44 &  &   20 31 46.05  &  40 43 24.6 &  -     &  B0.5\,IV      &  0.426  &   0.227     \\
CPR02-A45 &  &   20 29 46.66  &  41 05 08.3 & 11.9   &  B0.5\,Vn      &  0.397  &   0.179     \\
CPR02-A46 &  &   20 31 00.19  &  40 49 49.7 & 11.3   &  O7\,Vf        &  0.362  &   0.190     \\
\hline
\end{longtable}
\end{center}

{\bf Notes}:
\vskip2mm

       ALS = Reed (1998, 2005);

       CP05 = Camer\'on \& Pasquali (2005);

       CPR02 = Camer\'on et al. (2002);

       MT91 = Massey \& Thompson (1991);

       SMV89 = Strai\v{z}ys et al. (1989);

       SKV93 = Strai\v{z}ys et al. (1993);

       LS02 = Laugalys \& Strai\v{z}ys (2002);

       LSV06-1 = Laugalys et al. (2006a); %NGC 6997

       LSV06-22 = Laugalys et al. (2006b, Table 2); % L935 cloud

       LSV07 = Laugalys et al. (2007);  % Collinder 428

       VI Cyg = numbers of the Cyg OB2 association stars from Johnson \&
Morgan (1954) and Morgan et al. (1954). VI Cyg is the former name of the
Cyg OB2 association.

       SMV89-248 = V2200 Cyg (B9p, $\alpha^2$ CVn type);

       VI Cyg 5 = V729 Cyg, EB;

       ALS\,11633, SMV89-159 and SMV89-188 -- spectral binaries;

       ALS\,17620, ALS\,17622 and SMV89-185 -- visual binaries.

The stars with spectral types marked by asterisks are classified in MK
system by C. J. Corbally:  one asterisk -- published in Strai\v{z}ys et
al.  (1999), two asterisks -- published in the present table.  Here are
notes on the spectra of individual stars (IS means the interstellar band
at 443 nm). Interstellar extinction values $A_V$ are from Laugalys \&
Straizys (2002).

LS02-40: IS band slight;

LS02-209: IS band moderate, $A_V$ = 2.9;

LS02-217: IS band slight, He slightly strong, $A_V$ = 2.3;

LS02-222:  very strong IS band, quite strong IS Ca K, $A_V$ = 4.2;

LS02-229:  sharp hydrogen line cores, moderate IS band, Hg-Mn, Cr-Sr;

LS02-230:  emission in H and He, with normal decrement, spectral class
is an estimate from H wings and He\,I D-series, IS band moderate, $A_V$
= 2.5;

LS02-267: IS band strong, $A_V$ = 2.2;

LS02-401: IS band slight, IS Ca K quite strong, $A_V$ = 1.9;

LS02-463: all He lines are filled in by emission, $A_V$ = 1.9;

LS02-476: IS band moderate; $A_V$ = 2.7;

LS02-537: IS band moderate, $A_V$ = 2.2;

LS02-584: IS band moderate, IS Ca K line strong, $A_V$ = 3.9;

LS02-608: Balmer line decrement weak, IS band slight, IS Ca K line
moderate, $A_V$ = 1.8;

LS02-691: noisy spectrum, IS band strong, $A_V$ = 5.2.

\vskip4mm

%\newpage

\noindent Cyg OB2 association located behind the Great Cygnus Rift,
4\degr\ from the NAP nebulae.  Investigations of the extinction law in
Cygnus discussed in our earlier paper (Strai\v{z}ys et al. 1999) do not
show any significant differences in extinction properties between
various directions in Cygnus.  Table 1 lists 95 OB-type stars from the
NAP region and 98 O--B1 stars from the Cyg OB2 association.  We list
only those Cyg OB2 stars which were used for plotting the reddening
line.  They include 15 brightest stars from Johnson \& Morgan (1954) and
Morgan et al.  (1954), 45 stars from Massey \& Thompson (1991) and 42
stars from Comer\'on et al.  (2002).  The last list contains stars
having `featureless' infrared spectra and considered as the candidate
O-type stars.  Hanson (2003) and Negueruela et al. (2008) have
classified
27 of them in MK and
confirmed that they indeed are O--B0 type stars.  The stars with blended
images have been excluded.  We also excluded two stars from the Massey
\& Thompson list (575 and 793) which show a considerable deviation from
the reddening line of other O--B1 stars.  The reddest star in the NAP
region is the CP05\,4 star with spectral type O5  determined by
Comer\'on \& Pasquali (2005).  The $J$--$H$ and $H$--$K_s$ color indices
given in the table were calculated from the 2MASS {\it J, H} and $K_s$
data.

\sectionb{3}{INTRINSIC COLOR INDICES}

Despite a wide use of the 2MASS photometric system, intrinsic color
indices ($J$--$H$)$_0$ and ($H$--$K_s$)$_0$ of stars of different
spectral and luminosity classes are unknown.  Usually they are being
obtained by transformation from the Koornneef (1983) or Bessell \& Brett
(1988) tabulations with the Carpenter (2001) equations.  Since these
transformation equations for O and B stars are rather uncertain, we
decided to determine their intrinsic color indices directly in the 2MASS
system by dereddening relatively bright stars with small interstellar
reddening.

For determining the intrinsic color indices for O- and B-type stars we
took some little reddened stars listed in Table 2. The three O-stars are
the least reddened field stars.  The B5--B6 and B8--B9 stars were
selected in the vicinity of the NAP nebulae from our Table 1 and from
the Fehrenbach et al.  (1961) catalog of stars in the Kapteyn Selected
Area 40.  For each star color excesses $E_{B-V}$ were transformed to
$E_{J-H}$ and $E_{H-K}$ by the equations given by Bessell \& Brett
(1988).  Since the reddenings are small,
$E_{H-K}$\,$\approx$\,$E_{H-K_s}$.  After that color indices were
dereddened for all stars individually taking differences of the observed
color indices and the corresponding color excesses:
$$
(J - H)_0 = (J - H) - E_{J-H}~, \eqno(1) $$  $$
(H - K_s)_0 = (H - K_s) - E_{H-K_s}. \eqno(2)
$$
Then dereddened color indices were averaged to obtain the intrinsic
color indices ($J$--$H$)$_0$ and ($H$--$K_s$)$_0$ for O8, B5.5 and B8.5
stars listed in Table 3.

\sectionb{4}{EQUATIONS OF THE REDDENING LINES}

Table 1 stars were divided into three spectral groups:  O--B1, B2--B6
and B7--B9.5, neglecting their luminosity classes.  However, we excluded
all emission-line B-stars which exhibit excesses of $H$--$K_s$ at
constant $J$--$H$.  For each spectral group we have plotted the $J$--$H$
vs.  $H$--$K_s$ diagram shown in Figures 1--3.  Figure 1 shows that
O--B1 type stars in the NAP region and in the Cyg OB2 association
exhibit the same slope of the reddening line.  The CP05\,4 star at
$H$--$K_s$ = 0.47 (the uppermost dot) lies also together with the
association stars.  Two Cyg OB2 stars, No.\,5 (O7e) and No.\,12
(B5\,Iab), deviate downwards from the reddening line considerably,
imitating the presence of circumstellar thermal emission in the dust or
electron free-free transitions.  Peculiarities of star No.\,12 were
widely discussed by Massey \& Thompson (1991); they find H$\beta$ line
in emission.  In the direction of these two stars condensations of CO
have been discovered (Scappini et al. 2002; Casu et al. 2005).  These
two stars were rejected from the reddening line solutions.  After the
listed rejections, we have 118 O--B1 stars, 29 B2--B6 stars and 46
B7--B9.5 stars.

%%%%%%%%%%%%%%%%%%%%%%%%%%%%%%%%%%%  TABLE 2

\begin{table}[!th]
\begin{center}
\vbox{\footnotesize\tabcolsep=6pt
\begin{tabular}{l@{\hskip9mm}lccc}
\multicolumn{5}{c}{\parbox{85mm}{\baselineskip=9pt
{\smallbf Table 2.}{\small\ Color excesses of little reddened O--B
stars used in the determination of intrinsic color indices.\lstrut}}}\\
 [+4pt]
\tablerule
Name & \kern-10pt Spectral type & $E_{B-V}$ & $E_{J-H}$ & $E_{H-K_s}$ \\
\tablerule
\noalign{\vskip2mm}
O-type stars  &          &            &            &          \\
\noalign{\vskip2mm}
S Mon     &    O7\,Ve    &      0.09  &     0.033  &   0.017  \\
68 Cyg    &    O8e       &      0.26  &     0.096  &   0.050  \\
10 Lac    &    O9\,V     &      0.11  &     0.041  &   0.021  \\
\noalign{\vskip2mm}
B5--B6 stars &           &            &            &          \\
\noalign{\vskip2mm}
BD+45 3242  &  B5\,V     &      0.19  &     0.073  &   0.036  \\
BD+45 3279  &  B6\,V     &      0.25  &     0.092  &   0.048  \\
HD 198915   &  B5\,V     &      0.10  &     0.037  &   0.019  \\
BD+44 3579  &  B5\,V     &      0.30  &     0.111  &   0.057  \\
BD+46 3141  &  B5\,V     &      0.06  &     0.022  &   0.011  \\
\noalign{\vskip2mm}
B8--B9 stars &           &            &            &          \\
\noalign{\vskip2mm}
HD 197374   &  B9\,V     &      0.00  &     0.000  &   0.000  \\
HD 197391   &  B8\,V     &      0.15  &     0.055  &   0.028  \\
HD 199121   &  B8\,V     &      0.00  &     0.000  &   0.000  \\
HD 199417   &  B9\,V     &      0.10  &     0.037  &   0.019  \\
BD+45 3247  &  B9\,V     &      0.12  &     0.044  &   0.023  \\
BD+43 3701  &  B9\,V     &      0.08  &     0.030  &   0.015  \\
BD+45 3256  &  B8\,V     &      0.09  &     0.033  &   0.017  \\
BD+45 3264  &  B8\,III   &      0.14  &     0.052  &   0.027  \\
\tablerule
\end{tabular}
}
\end{center}
\end{table}

%%%%%%%%%%%%%%%%%%%%%%%%%%%%%%%%%%%  TABLE 3

\begin{table}[!th]
\begin{center}
\vbox{\footnotesize\tabcolsep=8pt
\begin{tabular}{llccc}
\multicolumn{4}{c}{\parbox{60mm}{\baselineskip=9pt
{\smallbf Table 3.}{\small\ Intrinsic color indices of stars in the
2MASS system.\lstrut}}} \\ [+4pt]
\hline \hstrut
Color index      &    ~~O8    &    B5.5  &   B8.5     \\
\hline \hstrut
($J$--$H$)$_0$   &  --0.164 &  --0.081 &   --0.062  \\
($H$--$K_s$)$_0$ &  --0.058 &  --0.035 &   --0.009  \\
\hline
\end{tabular}
}
\end{center}
\end{table}

%%%%%%%%%%%%%%%%%%%%%%%%%%%%%%%%%%%% FIGURE 1

\begin{figure}[!t]
\vbox{
\centerline{\psfig{figure=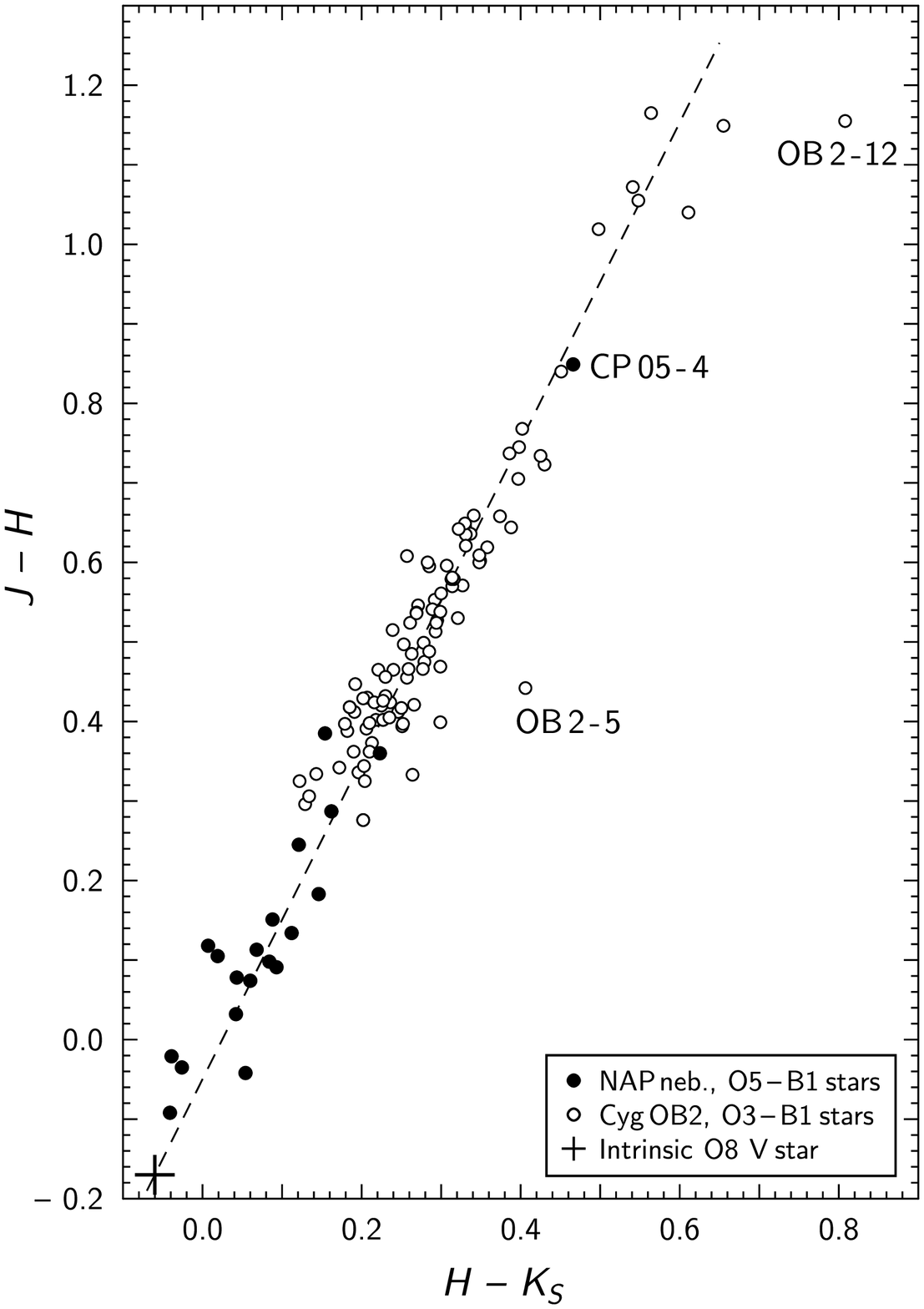,width=90truemm,angle=0,clip=}}
\vspace{.5mm}
\captionb{1}{Interstellar reddening line for O--B1 stars in the NAP
nebulae region and the Cyg OB2 association. The broken line is the
least-square solution for all 118 stars with the fixed intrinsic colors
$J$--$H$ = --\,0.17, $H$--$K_s$ = --\,0.06. The Cyg OB2 stars Nos. 5
and 12 are rejected.}
}
\end{figure}

%%%%%%%%%%%%%%%%%%%%%%%%%  FIGURES 2 and 3

\begin{figure}[!t]
\vbox{
\parbox[t]{61mm}{\psfig{figure=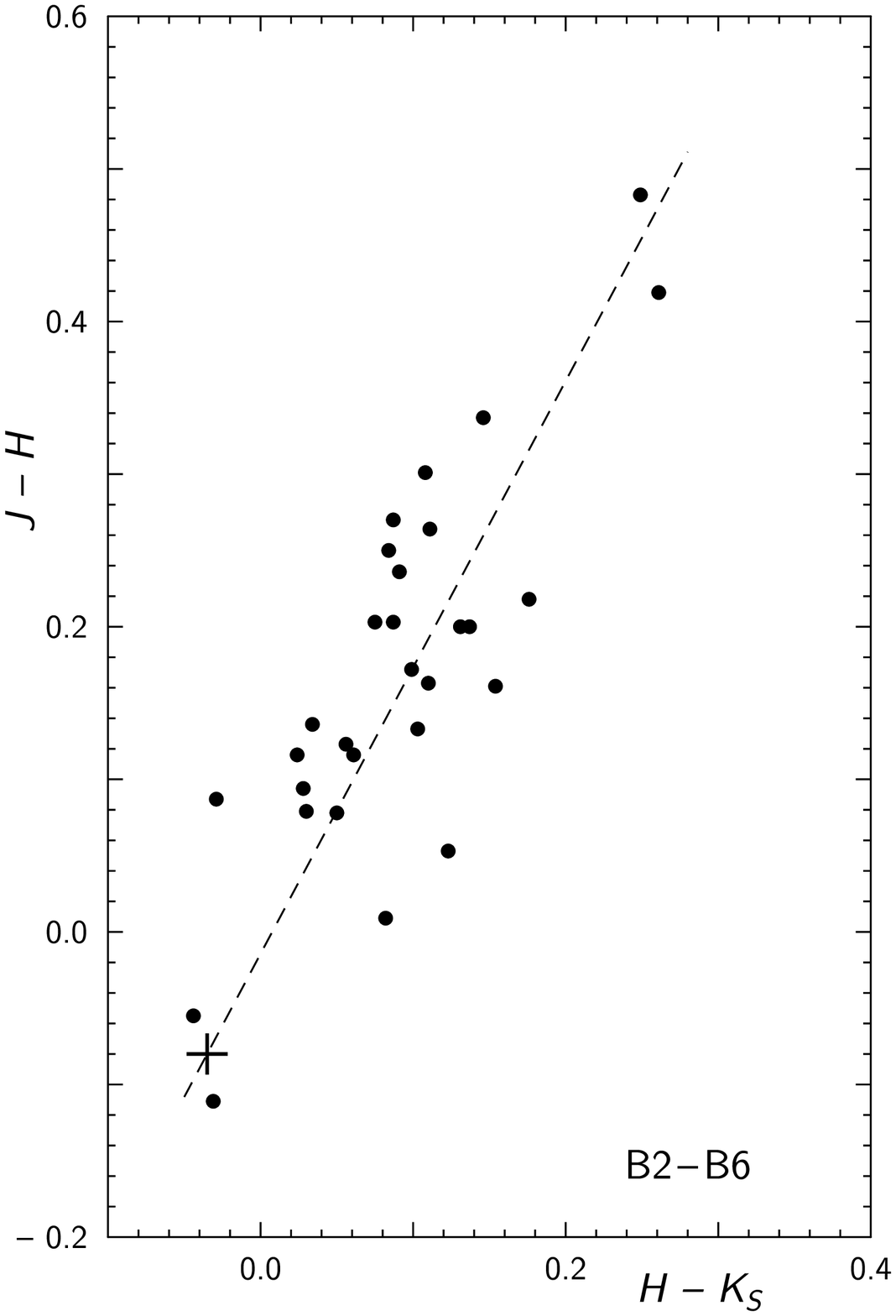,angle=0,width=61truemm,clip=}}
\hskip3mm
\parbox[t]{61mm}{\psfig{figure=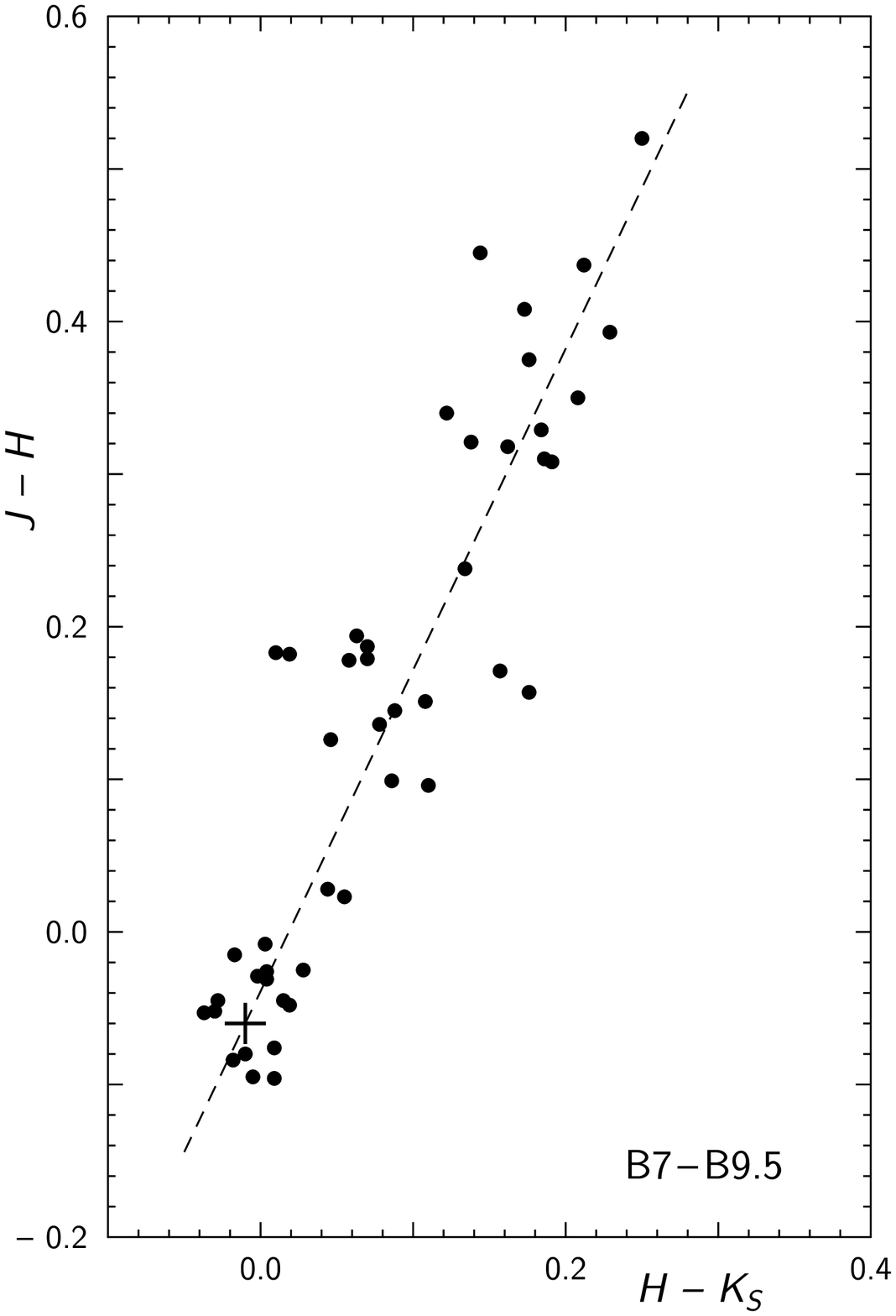,angle=0,width=61truemm,clip=}}
\vskip2mm

\captionb{2}{Interstellar reddening line for B2--B6 stars in the NAP
nebulae region. The broken line is the least-square solution for 29
stars with the fixed intrinsic colors $J$--$H$ = --\,0.08, $H$--$K_s$ =
--\,0.035.}
\vskip1mm
\captionb{3}{Interstellar reddening line for B7--B9.5 stars in the NAP
nebulae region. The broken line is the least-square solution for 46
stars with the fixed intrinsic colors $J$--$H$ = --\,0.06, $H$--$K_s$ =
--\,0.01.}
}
\vskip5mm

\begin{center}
\vbox{\footnotesize\tabcolsep=5pt
\centerline{\baselineskip=9pt
{\smallbf Table 4.}{\small\ Red clump giants in the M\,67
cluster.\lstrut}}
\begin{tabular}{lcccc}
\hline \hstrut
Star       &    $V$  & $B$--$V$ &  $J$--$H$ & $H$--$K_s$  \\
\hline \hstrut
MMJ 6485   &   10.48 &  1.11  &    0.485  &    0.133  \\
MMJ 6492   &   10.59 &  1.12  &    0.528  &    0.146  \\
MMJ 6494   &   10.48 &  1.10  &    0.506  &    0.153  \\
MMJ 6503   &   10.55 &  1.12  &    0.494  &    0.114  \\
MMJ 6506   &   10.58 &  1.10  &    0.504  &    0.118  \\
MMJ 6512   &   10.55 &  1.10  &    0.513  &    0.125  \\
MMJ 6516   &   10.47 &  1.12  &    0.485  &    0.164  \\
\hline
\noalign{\vskip2mm}
           &         & Average  &  0.502  &    0.136  \\
\end{tabular}
}
\end{center}

\end{figure}

The least-square solutions for Figures 1--3 have been made with the
fixed intrinsic positions of O--B1, B2--B6 and
B7--B9.5, respectively. The following equations were obtained:
$$ J-H = 2.004 (\pm 0.016) (H-K_s) - 0.050~,  \eqno(3)$$
$$ J-H = 1.876 (\pm 0.105) (H-K_s) - 0.014~, \eqno(4)$$
$$ J-H = 2.106 (\pm 0.094) (H-K_s) - 0.039~.  \eqno(5)$$
These equations show the slope of the reddening line,
$E_{J-H}/E_{H-K_s}$, for O--B stars is between 1.9 and 2.1.  Probably,
the average value 2.0 can be accepted for future analysis of the
distribution of reddened stars in the 2MASS two-color diagram.
Zero-points of the equations mean the points on the $J$--$H$ axis
at which the reddening lines cross the line $H$--$K_s$ = 0.0. They are
the same as the values of the interstellar reddening-free $Q_{JHKs}$
parameters:
$$
Q_{JHK_s} = (J - H) - E_{J-H}\,/\,E_{H-K_s} (H-K_s)~. \eqno(6)
$$
Equations (3), (4) and (5) show that the $Q_{JHK_s}$ values for the
three spectral classes are --0.050, --0.014 and --0.039. The maximum
absolute deviations of individual values from the mean  are  0.015, and
this means that all O--B stars lie practically on one line which
coincides with the reddening line (Figure 4). Thus, we may solve the
least square equation using all 193 stars together (with the intrinsic
position of an O8-type star):
$$
J-H = 2.024 (\pm 0.018) (H-K_s) - 0.048~. \eqno(7)
$$

%%%%%%%%%%%%%%%%%%%%%%%%%%%%%%  FIGURE 4

\begin{figure}[!t]
\vbox{
\centerline{\psfig{figure=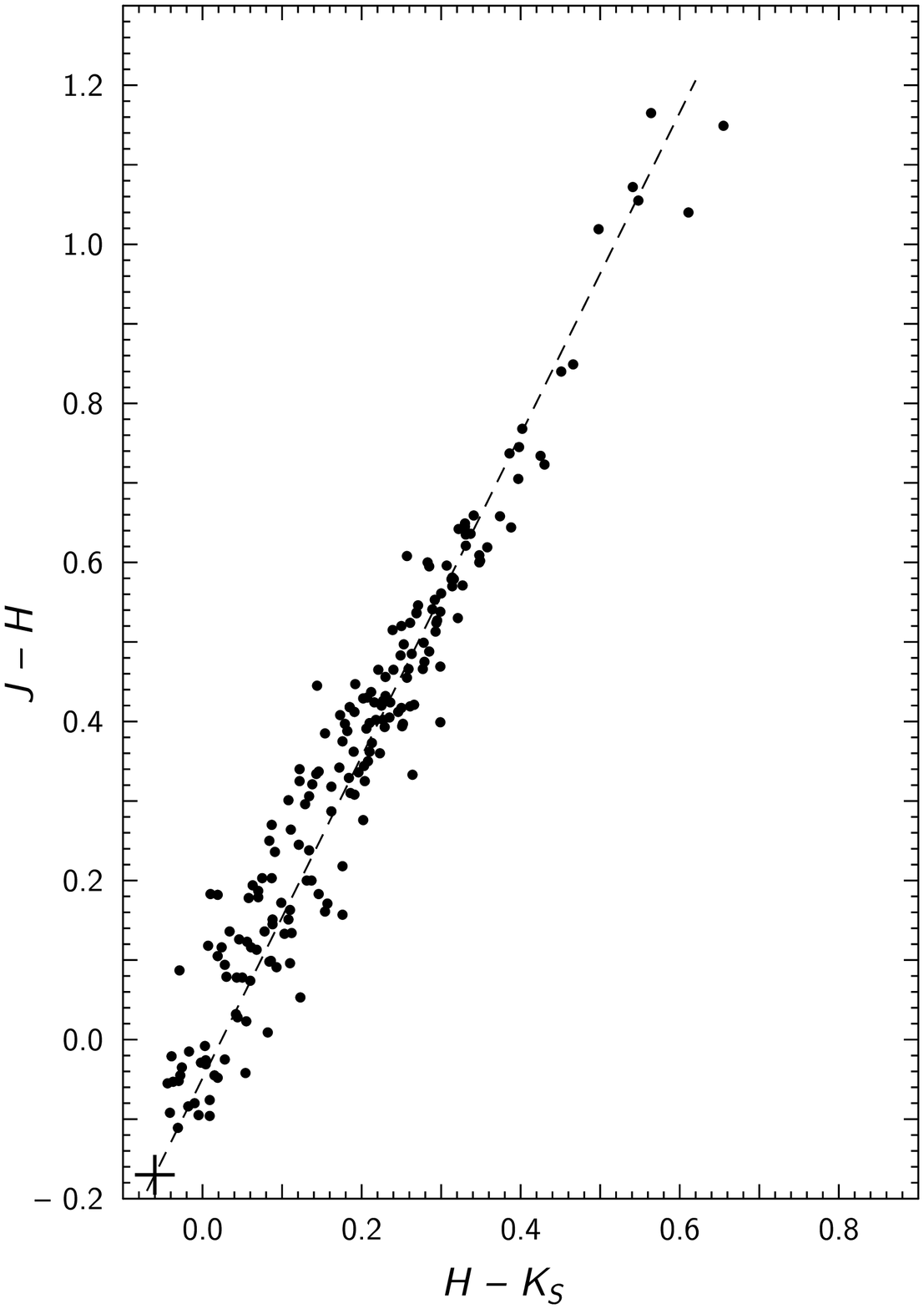,width=90truemm,angle=0,clip=}}
\vspace{.5mm}
\captionb{4}{Interstellar reddening line  in the
NAP nebulae region and the Cyg OB2 association for O--B9.5 stars
together. The broken line is the
least-square solution for all 193 stars with the fixed intrinsic colors
$J$--$H$ = --\,0.17, $H$--$K_s$ = --\,0.06. }
}
\end{figure}

\sectionb{5}{THE REDDENING LINE OF RED GIANTS}

%%%%%%%%%%%%%%%%%%%%%%%%%%%%%%  FIGURE 5

\begin{figure}[!t]
\vbox{
\centerline{\psfig{figure=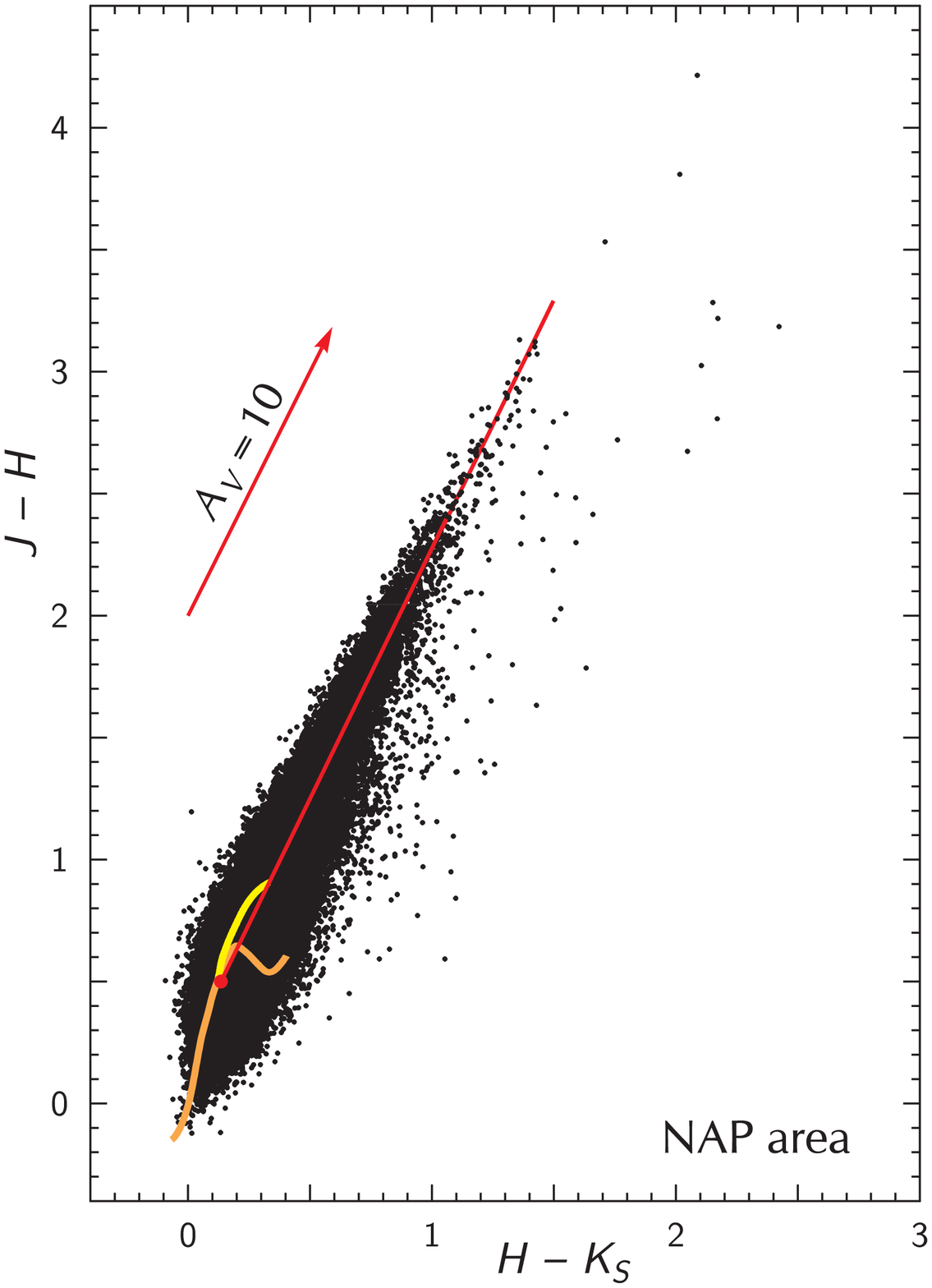,width=100mm,angle=0,clip=}}
\vspace{.5mm}
\captionb{5}{Two-color diagram for the NAP nebulae area of
$3\degr\times 3\degr$ size. The orange and yellow curves are
intrinsic sequences of luminosity V and III stars. The red dot is the
intrinsic position of the red clump giants, whose interstellar
reddening is shown by the straight red line.
 The red line with an arrow is the reddening
vector, its slope is $E_{J-H}\,/\,E_{H-K_s}$ = 2.0 and its length
corresponds to the extinction $A_V$ = 10 mag.}
}
\end{figure}

%%%%%%%%%%%%%%%%%%%%%%%%%%%%%%  FIGURE 6

\begin{figure}[!t]
\vbox{
\centerline{\psfig{figure=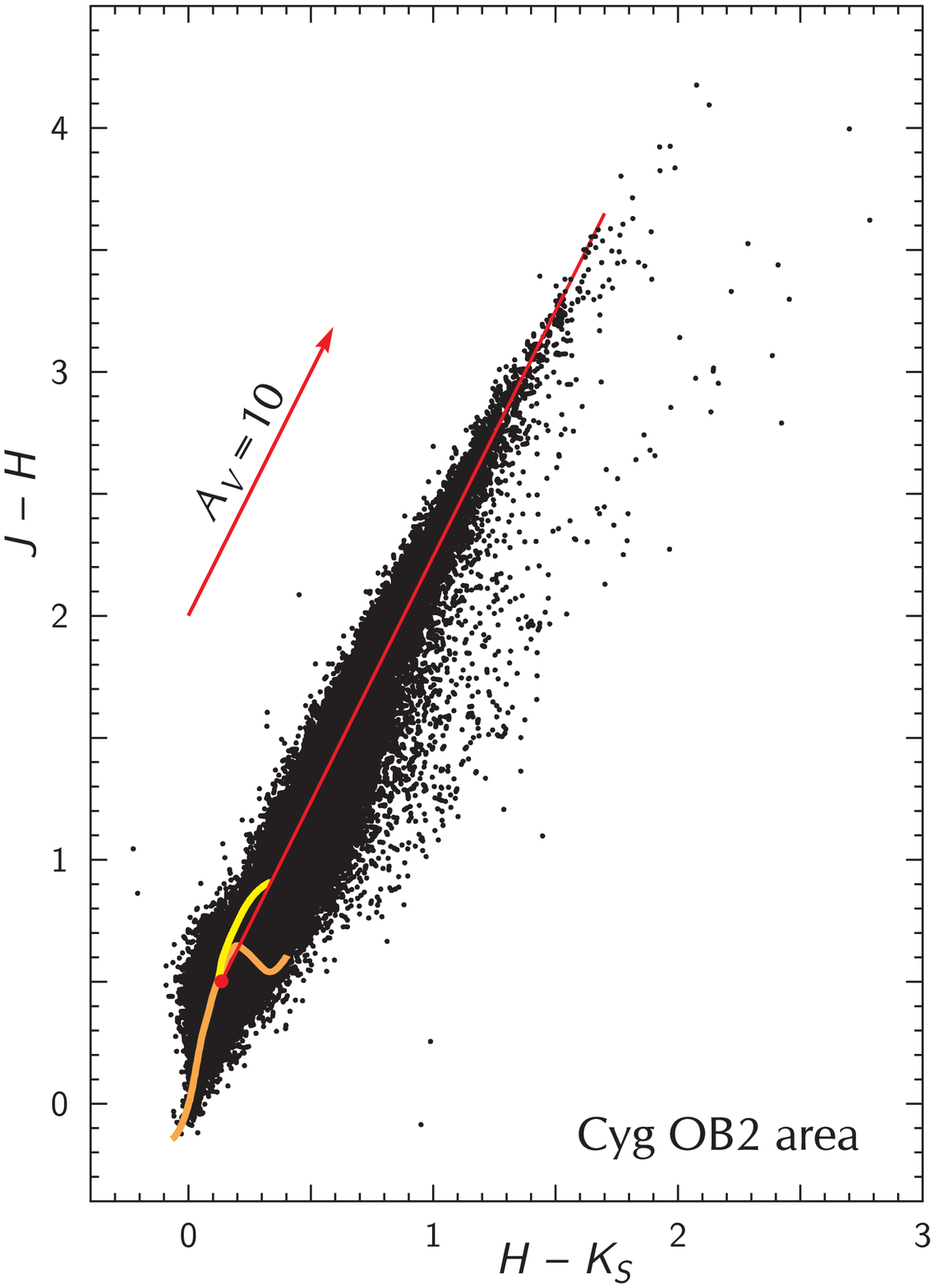,width=100mm,angle=0,clip=}}
\vspace{.5mm}
\captionb{6}{The same as in Figure 5 but for the Cyg OB2 association
area of
$3\degr\times 3\degr$ size and the center $20^{\rm h}\,33^{\rm m}$,
$+41\degr\,20\arcmin$ (J2000).}
}
\end{figure}

We have one more possibility to find the reddening line slope in the
$J$--$H$ vs.  $H$--$K_s$ diagram.  In Figure 5 we show the plot of this
diagram for 99\,000 stars in the 3\degr\,$\times$\,3\degr\ area with the
center given in Section 2. These stars were selected from the 2MASS
database with an error limit of $<$\,0.05 mag for the three magnitudes.
The stars form a comet-like crowding in which we show approximate
intrinsic positions of the main sequence and K--M giants.  The
interstellar reddening vector is shown, its slope is 2.0 and its length
corresponds to the extinction $A_V$ = 10 mag.  The tail of the `comet'
is composed of normal reddened stars, mostly of red clump giants of
early K subclasses (the discussion see in L\'opez-Corredoira et al.
2002).  This means that the stars at the upper end of the tail are
reddened K-type red clump giants with $A_V$\,$\approx$\,30 mag. The
comparison of their colors with the intrinsic colors of red giants may
be used to estimate the slope of the reddening line. However, we need to
know the intrinsic position of red clump giants in the $J$--$H$ vs.
$H$--$K_s$ diagram.

For determining the intrinsic position of red clump giants we used
the old open cluster M\,67 which is very suitable, as the
cluster is practically unreddened and its stars have solar chemical
composition. Seven clump stars with $V$ at 10.5 and $B$--$V$ at 1.1 were
selected from the Montgomery et al. (1993) catalog and are listed in
Table 4.

The straight line connecting the center of red clump stars with the end
of the `comet' tail at $J$--$H$ = 3.1 and $H$--$K_s$ = 1.4 has the slope
2.06. This value in good agreement with the values obtained for O-
and B-stars in Section 4.

In Figure 6 we show a similar diagram for 66\,000 stars selected in the
association Cyg OB2 area.  The area is of $3\degr\times 3\degr$ size and
its center is at $20^{\rm h}\,33^{\rm m}$, $+41\degr\,20\arcmin$
(J2000).  The straight red line joins the intrinsic position of the red
clump giants and the upper end of the `comet' tail at $J$--$H$ = 3.65
and $H$--$K_s$ = 1.56.  The slope of the reddening line, drawn by eye,
is $E_{J-H}\,/\,E_{H-K_s}$ is 2.02, in perfect agreement with the NAP
area.

\sectionb{6}{DISCUSSION AND CONCLUSIONS}

The ratios $E_{J-H}/E_{H-K_s}$ determined so far in the 2MASS system in
various Milky Way areas exhibit quite a wide range of values, between
1.6 and 2.1, the average value being close to 1.8.  Our slopes of
reddening lines in the NAP area and the Cyg OB2 association area both
for O--B stars and K giants are near the upper limit of this range.
Only for Globule 2 in the Coalsack Racca et al.  (2002) in the CIT
system obtained $E_{J-H}/E_{H-K}$ = 2.08.  If the ratio of colors in
both systems given by Carpenter (2001) is valid also for color excesses,
the ratio $E_{J-H}/E_{H-K_s}$ in the 2MASS system should be 1.05 times
larger, i.e., in the Coalsack it can be close to 2.2.

It is obvious that the variations of color-excess ratios in the infrared
are related to sizes and compositions of dust grains.  Most authors
agree that between $\sim$\,0.9 $\mu$m and $\sim$\,5 $\mu$m the
extinction curve can be approximated by a power law, $A_{\lambda}
\propto \lambda^{- \beta}$.  The most widely used is the value $\beta$ =
1.8, representing the so-called `universal' extinction curve in the
infrared (Martin \& Whittet 1990).  However, in the last decade it was
realized that $\beta$ is not universal but has different values between
1.6 and 1.8 (see Draine 2003; Indebetouw et al. 2005; Flaherty et al.
2005; Froebrich \& Burgo 2006; Froebrich et al. 2007).  Since
$$
\ln (A_1 / A_2) = \beta \ln (\lambda_2 /\lambda_1), \eqno(8)
$$
for the effective
wavelengths of the passbands $J$, $H$ and $K_s$ we can calculate the
approximate relation between $\beta$ and the ratio of color excesses.
Convolving the response functions given by Cutri et al.  (2006) and
Skrutskie et al.  (2006) with spectral energy distributions of Kurucz
models we obtain that $\lambda_{\rm eff}$ values for different
temperatures are not very different, thus we took the values of 1.24,
1.64 and 2.14 $\mu$m corresponding to solar-type stars.  In this case
$$
\beta = 2.045 (E_{J-H}/E_{H-K_s}) - 1.722. \eqno(9)
$$
For the NAP area, where $E_{J-H}/E_{H-K_s}$ = 2.0, we obtain $\beta$ =
2.37.  This value of $\beta$ gives a relatively steep interstellar
extinction curve in the 1--2 $\mu$m range of wavelengths.
A similar value of $\beta$ was recently obtained by Larson \& Whittet
(2005) for high Galactic latitude clouds. Whittet (2008) estimates that
such a value of $\beta$ suggests smaller than average grain sizes,
compared with the `typical' value of $\beta$ = 1.8. The curvature of the
near IR segment of the extinction curve is a reflection of the fact that
even the larger grains have sizes which are smaller than $\lambda$. In
the small particle limit one would expect $\beta$ = 4.0 (Rayleigh
scattering). On the other hand, for larger grains (e.g., with dimensions
$\sim$\,2 $\mu$m) one would expect $\beta$\,$\approx$\,1.0. It is not
really possible to estimate average grain sizes from $\beta$ but it
should certainly follow the trend: larger $\beta$, smaller grains.

On the other hand, according to our earlier investigations, the dust in
L\,935 and the surrounding NAP area exhibits other peculiarities.
Earlier we have obtained that the extinction law in the vicinity of NAP
exhibits a smaller `knee' in the blue part of the spectrum (Strai\v{z}ys
et al. 1999) which is also consistent with smaller grains responsible
for the extinction in the range of wavelengths covered by the $B$ and
$V$ passbands.

To summarize, the following results of the present investigation may be
listed.

1. A list of 95 O- and B-type stars with MK classifications,
supplemented by the 2MASS $J$--$H$ and $H$--$K_s$ color indices, is
compiled in the $3\degr\,\times\,3\degr$ area covering the North America
and Pelican nebulae and including the L\,935 dust cloud.  For 37 stars
spectroscopic MK types and for 40 stars photometric types are
determined by the authors. The list is supplemented by 98 O--B1 type
stars from the Cyg OB2 association.

2. Intrinsic color indices ($J$--$H$)$_0$ and ($H$--$K_s$)$_0$  are
determined for spectral classes O8, B5.5 and B8.5 by dereddening
bright stars with small interstellar extinction.

3. Interstellar reddening lines are calculated for stars of the three
spectral groups:  O--B1, B2--B6 and B7--B9.5.  The slopes of the
reddening lines, 2.00, 1.88 and 2.10, are obtained for the three groups.

4. The mean intrinsic colors $J$--$H$ and $H$--$K_s$ of seven red clump
giants of spectral types G8--K2\,III in the open cluster M\,67 are
determined.  For areas of both the NAP nebulae and the Cyg OB2
association, joining the positions of the unreddened clump giants and
the most heavily reddened stars in the $J$--$H$ vs.  $H$--$K_s$ diagram,
we obtain the reddening line slope of 2.06 and 2.02, respectively, which
are in a good agreement with the slopes for O--B stars.

5. The mean ratio of color excesses $E_{J-H}/E_{H-K_s}$ = 2.0
may be recommended for the North America and Pelican nebulae region, as
well as for the Cyg OB2 association. This value is somewhat larger than
the ratios which are usually in use in the Cygnus direction.

\thanks{We are thankful to the Steward Observatory for allocation  of
the observing time. The use of the 2MASS, Simbad and Gator databases is
acknowledged. V.S. is thankful to D.\,C.\,B. Whittet and B. T. Draine
for important comments.}

\References

\refb Bessell M. S., Brett J. M. 1988, PASP, 100, 1134

\refb Cambr\'esy L., Beichman C. A., Jarrett T. H., Cutri R. M. 2002,
AJ, 123, 2559

\refb Carpenter J. M. 2001, AJ, 121, 2851

\refb Casu S., Scappini F., Cecchi-Pestellini C., Olberg M. 2005, MNRAS,
359, 73  % Cyg OB2 No.5 and No.12

\refb Comer\'on F., Pasquali A. 2005, A\&A, 430, 541

\refb Comer\'on F., Pasquali A., Rodighiero G., Stanishev V. et al.
2002, A\&A, 389, 874

\refb Cutri R. M., Skrutskie M. F., Van Dyk S., Beichman C. A. et al.
2006, Eplanatory Supplement to the 2MASS All Sky Data Release and
Extended Mission Products, \\
http://www.ipac.caltech.edu/2mass/releases/allsky/doc/explsup.html

\refb Djupvik A. A., Andr\'e Ph., Bontemps S., Motte F. et al. 2006,
A\&A, 458, 789

%\refb Drimmel R., Cabrera-Lavers A., L\'opez-Corredoira M. 2003, A\&A,
%409, 205

\refb Draine B. T. 2003, {\it Interstellar Dust Grains}, ARA\&A, 41, 241

\refb Fehrenbach Ch., Petit M., Cruvellier G., Peyrin Y. 1961, J. des
Observateurs, 44, 233

\refb Fitzpatrick E. L., Massa D. 2005, AJ, 130, 1127

\refb Fitzpatrick E. L., Massa D. 2007, ApJ, 663, 320

\refb Flaherty K. M., Pipher J. L., Megeath S. T., Winston E. M. 2007,
ApJ, 663, 1069

\refb Froebrich D., del Burgo C. 2006, MNRAS, 369, 1901

\refb Froebrich D., Murphy G. C., Smith M. D., Walsh J., del Burgo C.
2007, MNRAS, 378, 1447

\refb Hanson M. M. 2003, ApJ, 597, 957

\refb He L., Whittet D.\,C.\,B., Kilkenny D., Spencer Jones J. H. 1995,
ApJS, 101, 335

\refb Indebetouw R., Mathis J. S., Babler B. L., Meade M. R. et al.
2005, ApJ, 619, 931

\refb Johnson H. L., Morgan W. W. 1954, ApJ, 119, 344

\refb Koornneef J. 1983, A\&A, 128, 84

\refb Laugalys V., Strai\v zys V. 2002, Baltic Astronomy, 11, 205

\refb Laugalys V., Strai\v zys V., Vrba F. J., Boyle R. P., Philip
A.\,G.\,D., Kazlauskas A. 2006a, Baltic Astronomy, 15, 327
% NGC 6997

\refb Laugalys V., Strai\v zys V., Vrba F. J., Boyle R. P., Philip
A.\,G.\,D., Kazlauskas A. 2006b, Baltic Astronomy, 15, 483

\refb Laugalys V., Strai\v zys V., Vrba F. J., \v{C}ernis K.,
Kazlauskas A., Boyle R. P., Philip A.\,G.\,D. 2007, Baltic Astronomy,
16, 349           % Collinder 428

\refb Lombardi M., Alves J., Lada C. J. 2006, A\&A, 454, 781

\refb L\'opez-Corredoira M., Cabrera-Lavers A., Garz\'on F., Hammersley
P. L. 2002, A\&A, 394, 883

\refb Massey P., Thompson A. B. 1991, AJ, 101, 1408

%\refb Marshall D. J., Robin A. C., Reyl\'e C., Schultheis M., Picaud S.
%2005, A\&A, 453, 635

\refb Montgomery K. A., Marschall L. A., Janes K. A. 1993, AJ, 106, 181

\refb Morgan W. W., Johnson H. L., Roman N. G. 1954, PASP, 66, 85

\refb Naoi T., Tamura M., Nakajima Y., Nagata T. et al. 2006, ApJ, 640,
373

\refb Negueruela I., Marco A., Herrero A., Clark J. S. 2008, A\&A, 487,
575

\refb Nishiyama S., Nagata T., Kusakabe N., Matsunaga N. et al. 2006,
ApJ, 638, 839

\refb Racca G., G\'omez M., Kenyon S. J. 2002, AJ, 124, 2178

\refb Reed B. C. 1998, ApJS, 115, 271

\refb Reed B. C. 2005, AJ, 130, 1652

\refb Rieke G. H., Lebofsky M. J. 1985, ApJ,  288, 618

\refb Rom\'an-Z\'uniga C. G., Lada C. J., Muench A., Alves J. F. 2007,
ApJ, 664, 357

\refb Scappini F., Casu S., Cecchi-Pestellini C., Olberg M. 2002, MNRAS,
337, 495

\refb Skrutskie M. F., Cutri R. M., Stiening R., Weinberg M. D. et al.
2006, AJ, 131, 1163

%\refb Strai\v zys V. 1992, {\it Multicolor Stellar Photometry}, Pachart
%Publishing House,\\ Tucson, Arizona

\refb Strai\v{z}ys V., Corbally C. J., Laugalys V. 1999, Baltic
Astronomy, 8, 355

%\refb Strai\v zys V., Goldberg E. P., Mei\v stas E., Vansevi\v cius V.
%1989b, A\&A, 222, 82

\refb Strai\v zys V., Kazlauskas A., Vansevi\v cius V., \v Cernis K.
1993, Baltic Astronomy, 2, 171

\refb Strai\v zys V., Mei\v stas E., Vansevi\v cius V., Goldberg E. P.
1989, Bull. Vilnius Obs., No. 83, 3

\refb Whittet D.\,C.\,B. 2008, personal communication

\end{document}